\documentclass[a4paper]{article}
\usepackage{amsmath,amssymb,amsthm}
\usepackage{latexsym}
\usepackage{epsfig}

\title{Interaction
of a Supernova Shock with the other Star in a Binary System}
\author{Ya.~N. Istomin,$^{1}$
\thanks{E-mail: istomin@td.lpi.ac.ru}
 F.~L. Soloviev$^{2}$\thanks{E-mail: soloviev@nyu.edu}\\
$^{1}$P.N.Lebedev Physical Institute,\\
Leninsky Prospect 53, Moscow 119991, Russia\\
$^{2}$Courant Institute of Mathematical Sciences,\\
251 Mercer Street, New York, NY 10012-1185, USA}

\begin{document}

\date{Released 2004 Xxxxx XX}

\maketitle

\label{firstpage}

\begin{abstract}
Interaction of a fast shock wave generated during a supernova
explosion with a magnetized star-companion of the supernova precursor
produces a current sheet. We consider the evolution of this current sheet
and show that a singularity (shock) is formed in finite
time within the ideal MHD framework.
Charged particles (electrons) are accelerated in the vicinity of the
singularity, and their distribution function has a plateau up to the
energies of the order of $10^4 mc^2$.
These fast particles radiate in the $\gamma$-range in the strong magnetic field
of the current sheet ($B\simeq 10^6 G$).
Radiation is concentrated within a narrow angle around the
current sheet, $\Delta\theta \simeq 3\cdot 10^{-4}$, and its spectrum
has the maximum at several hundreds of $keV$.
Presented calculations confirm the model of cosmological GRBs proposed
by Istomin \& Komberg (2002).
\end{abstract}

\section{Introduction}

Observations show that some gamma-ray bursts (GRBs) occur at cosmological
distances, and some of the bursts are related to the supernova explosions.
If the radiation of a cosmological GRB is isotropic,
its energy release is estimated to be $10^{53-54}$ erg at the
hundreds of keV range.
Massive supernovae emit comparable amounts of energy, mainly in the form
of neutrinos. Thus a question about the origin of GRBs' energy supply
arises. Connection between the GRB phenomenon and that of a
supernova seems to be supported by several facts.
First, the paper by~\cite{WW98} reports the temporal
and spatial concurrence of the ``compact'' supernova SN 1998 bw
($10^{45}$ erg, type SNIc) with GRB 980425, identified with a nearby galaxy
which has $Z=0.0085$.
Secondly, ``hills'' are observed in the light curves of some optical
transients several days after the GRB; they can be interpreted as a
contribution from the supernova burst simultaneous with the GRB.
Thirdly, narrow X-ray lines of ions characteristic of
supernova shells are detected in the spectrum of the afterglow~\cite{RW02},
with the velocity of the line-emitting material reaching $0.1c$.
Obviously, the ``energy crisis'' arises if we attempt to relate an isotropic
GRB to a supernova. One way to avoid this crisis is to suggest that GRBs
are associated with strongly collimated relativistic ejections radiating
within $\gamma$-range. This makes it possible to decrease by many orders the luminosity
from cosmological GRBs corresponding to the observed fluxes. \cite{P01}
proposed this connection between GRBs and strongly anisotropic supernova
bursts. Summarizing, we see that current understanding of the
nature of GRBs requires a combination of two conditions: a supernova
explosion (possibly of special type) and the formation of a narrow
(opening angle less than a degree) beam of relativistic particles radiating
within X- to $\gamma$-range. \cite{IK03} proposed a
model where a beam of relativistic particles is formed by
the interaction of a fast shock wave generated during a supernova explosion
with a magnetized star-companion (a neutron star or a white dwarf) of the
supernova precursor. The flow of the shock matter around the magnetosphere
of the star-companion forms an almost parallel magnetospheric tail.
Magnetic field in the tail becomes super strong (about $10^6$ G)
due to the cumulative effect. Virtually any star-companion interacting
with a supernova shock will produce such strong magnetic field,
due to amplification during the
explosive compression of conducting medium - the mechanism proposed by~\cite{Sakh}
to produce very strong magnetic fields. For the typical shock
parameters, the density $\rho_{sh}$ is of order $10^{-8} g/cm^3$, and
the velocity $V_{sh}$ is about $4\cdot 10^9 cm/sec$, the value of the amplified
magnetic field is $B=(4\pi\rho_{sh}V_{sh}^2)^{1/2}\simeq 10^6$ Gauss.
For a usual star like the sun (which has magnetic field of the order of $10G$),
the compression will be about $3\cdot 10^2$ times and the transversal size of the tail
will be equal to $R_\odot/3 \cdot 10^2\simeq 10^8 cm$. Hence, virtually any
explosion of a supernova which had a star-companion, will result in a highly magnetized
elongated plasma formation in the shadow of the star-companion.
The magnetic field changes its direction from one edge of the tail to the
other, which leads to the magnetic reconnection, magnetic energy
release, and acceleration of charged particles.
Similar processes take place in the magnetospheric tail of the Earth formed by
the flow of the solar wind.
In a strong magnetic field
these relativistically accelerated particles produce synchrotron radiation
concentrated in a narrow cone $\Delta\theta\simeq 3\cdot 10^{-4}$.
The frequency of supernova explosions has the order of $3\cdot 10^{-2}supernovaes/yr$
per a galaxy, which gives $(3\cdot 10^{-4})^2 \cdot 3\cdot 10^{-2} \cdot 10^{11} \simeq 300$ radiation bursts
per year, directed towards the Earth (the Universe has $\simeq 10^{11}$ galaxies).
All this may provide an insight into the nature of GRBs.
In this paper we provide an accurate quantitative theory of the
outlined phenomena, namely the evolution of the current sheet,
the acceleration of charged particles, and the formation of the radiation
spectrum.

\section[]{Stationary current sheet}

Magnetic field of a star is almost dipolar at the large distances from the
center of the star. Its direction changes from one
side of the sheet to the other (with respect to the axis of the tail).
Interaction of a dense plasma flow with this dipole magnetic field produces
a long current sheet.
A similar magnetospheric tail is formed by solar wind in the dipolar magnetic
field of the Earth. Such situation may also arise in a close
binary system of stars, consisting of a magnetized star and a massive
supernova precursor.
A shock wave from the exploding massive star interacts with the magnetized
star and forms a current sheet, resulting in a GRB.
Such model of GRBs was proposed in the papers of~\cite{IK03}.
Yet another example of the current sheet formation comes from a binary system of
neutron stars, one of which is a radio pulsar. Relativistic wind flow
of the radio pulsar around the other star forms a magnetic tail.
Discovery of a close binary system of neutron stars
containing two radio pulsars PSR J0737-3039 A,B was reported in the paper of~\cite{LB04}.
Observations of this system show that the wind flow of the more powerful
pulsar 'A' around the magnetosphere of the radio pulsar 'B' takes place
inside the light cylinder of the pulsar 'B'.
Thus, a long and almost parallel tail of magnetic field is formed in the
magnetosphere of 'B'.

To study the phenomena taking place in the magnetosphere of a magnetized star,
we assume the sheet to be flat and infinitely long.
Therefore, all characteristics depend only on one coordinate $z$,
which is perpendicular to the current sheet. Our approximation
is one-dimensional, which makes possible the solution of non-stationary problems.
The geometric structure of this problem is presented in Figure 1.
\begin{figure*}
\centering
\includegraphics{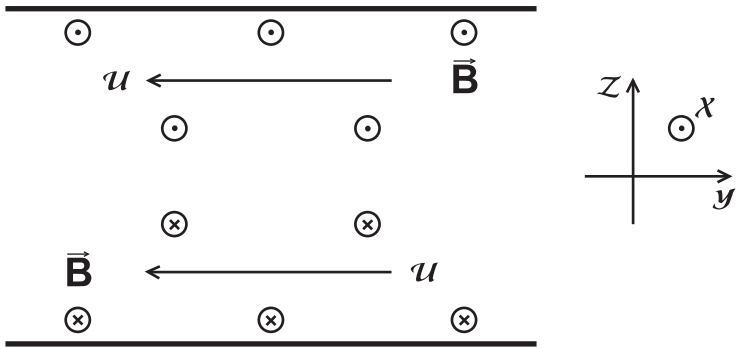}
\caption{Stationary current sheet}
\end{figure*}

Non-uniform magnetic field is parallel to the $x$ axis and the tail,
the $y$ axis is perpendicular to the $x$-$z$ plane.
Magnetic field depends linearly on the $z$ coordinate
\[
{\bf B} = \left(Kz, 0, 0\right),
\]
where $K$ is a constant determined by electric currents flowing in the $y$
direction. A particle with charge $q$ and mass $m$ in such non-uniform
magnetic field is subject to the gradient drift~\cite{S65}
\begin{equation}
\label{grad-dr}
u_y = -\frac{c p_\perp v_\perp}{2 m q K z^2}.
\end{equation}
Here $v_\perp$ and $p_\perp$ are the velocity and momentum components of the particle
perpendicular to the magnetic field.
Equation~(\ref{grad-dr}) is not applicable near the plane $z \simeq 0$.
The Larmor radius of the particle is greater than the inverse
gradient of the field there, so the drift approximation is not valid
when $K z^2 < cp_\perp/q$. Electric current $j_d$ produced by the drift
motion of charged particles with density $n$ is given by
\begin{equation}
{j_d}_y =-n \frac{<p_\perp v_\perp>}{2 m} \frac{c}{Kz^2},
\end{equation}
and is independent of their charge.
Brackets $<...>$ mean averaging over the distribution function of the
particles. In addition to the drift current resulting from the drift of
the particles in the non-uniform magnetic field, there also exists a
magnetization current ${\bf j}_m$ due to non-uniform distribution of
the magnetic moments of the particles, ${\bf j}_m = c \cdot curl {\bf M},
\,{\bf M}=n<{\bf \mu}>,\,{\bf \mu}=-{\bf B}(p_\perp v_\perp/2B^2)$. The sum
of the drift current and the magnetization current depends only on the
gradient of the transversal energy density of particles
\begin{equation}
j_y = -\frac{c}{B}\frac{d}{dz}\left(n\frac{<p_\perp v_\perp>}{2}\right).
\end{equation}
The total current ${\bf j}$ is formed by the electron and ion flows.
Assuming plasma to be quasineutral and consisting of
electrons and positively charged ions with mean charge $Ze$, we see that
the electric current is
\begin{equation}
j_y = -\frac{c}{B}\frac{d}{dz}\left[n_i(1+Z)\frac{<p_\perp v_\perp>}{2}
\right].
\end{equation}
On the other hand, we have
\[
{\bf j} = \frac{c}{4 \pi} curl {\bf B} = \frac{cK}{4\pi} {\bf e_y}.
\]
Finally,
\begin{equation}
\label{equilib}
\frac{B^2}{8\pi}+n_i(1+Z)<\frac{p_\perp v_\perp}{2}> = const =P_0.
\end{equation}
Equation~(\ref{equilib}) is virtually identical to the equilibrium condition
of an inhomogeneous plasma in a non-uniform magnetic field, when the
equilibrium is supported by external pressure $P_0$.

\section[]{Evolution of the current sheet}

Now we add an extra assumption that the current sheet is compressed by some external force
acting along the $z$ axis, and consider the evolution of such sheet.
We may assume that the evolution is fast enough and it can be described
by ideal magnetohydrodynamics (MHD). MHD equations for the one-dimensional
case (all parameters depend only on the $z$ coordinate) are
\begin{eqnarray}
&&\frac{\partial B}{\partial t} + v_{z} \frac{\partial B}{\partial z}
+ B \frac{\partial v_{z}}{\partial z} = 0, \nonumber \\
&&\frac{\partial v_{y}}{\partial t} + v_{z} \frac{\partial v_{y}}{\partial z}
= 0, \nonumber \\
&&\rho \left( \frac{\partial v_{z}}{\partial t} + v_{z} \frac{\partial v_{z}}
{\partial z} \right) + \frac{\partial P}{\partial z} + \frac{1}{4 \pi}
B \frac{\partial B}{\partial z} = 0, \label{maineq} \\
&&\frac{\partial \rho}{\partial t} + v_{z} \frac{\partial \rho}{\partial z}
 + \rho \frac{\partial v_{z}}{\partial z} = 0, \nonumber \\
&&\frac{\partial P}{\partial t} + v_{z} \frac{\partial P}{\partial z} +
\Gamma P \frac{\partial v_{z}}{\partial z} = 0.\nonumber
\end{eqnarray}
During the one-dimensional compression (or expansion) of the current sheet
magnetic field remains one-dimensional,
\[
{\bf B} = \left( B(z, t), 0, 0\right).
\]
In addition to the $y$ component of the velocity ${\bf v}$, the $z$
component becomes time-dependent as well
\[
{\bf v} = \left( 0, v_{y}(z, t), v_{z}(z, t)\right).
\]
The plasma density is $\rho = \rho(z, t)$, its pressure is
$P = P(z, t)$. We use the equation of state $P = C_{2} \rho ^ \Gamma$,
where $\Gamma$ is the adiabatic exponent.
From the first and the fourth equations of system~(\ref{maineq}),
we obtain
\[
\rho \frac{\partial B}{\partial t} + \rho v_{z} \frac{\partial B}{\partial z}
= -\rho B \frac{\partial v_{z}}{\partial z} =
B \frac{\partial \rho}{\partial t} + B v_{z} \frac{\partial \rho}
{\partial z},
\]
which leads to the following condition
\[
\rho \frac{dB}{dt} = B \frac{d\rho}{dt},
\]
which means that the magnetic field is ``frozen'' into the plasma.
Hence, we have $B \propto \rho$, so let $B = C_{1} \rho$.
Now, the third and the fourth equations of system~(\ref{maineq}) describe the evolution
of the velocity $v_z(z,t)$ and plasma density $\rho(z,t)$
\begin{eqnarray}
&&\rho \left( \frac{\partial v_{z}}{\partial t} + v_{z} \frac{\partial v_{z}}
{\partial z} \right) + C_{2} \Gamma \rho^{\Gamma - 1} \frac{\partial \rho}
{\partial z} + \frac{C^{2}_{1} \rho}{4 \pi} \frac{\partial \rho}{\partial z}
 = 0, \nonumber \\
&&\frac{\partial \rho}{\partial t} + v_{z} \frac{\partial \rho}{\partial z}
+ \rho \frac{\partial v_{z}}{\partial z} = 0. \label{twoeq}
\end{eqnarray}
The change of variables $(\rho, v_z) \to (t, z)$ reduces the system of non-linear
equations~(\ref{twoeq}) to the following linear system
\begin{eqnarray}
&&\frac{\partial z}{\partial \rho} - v \frac{\partial t}{\partial \rho} +
C_{2} \Gamma \rho^{\Gamma - 2} \frac{\partial t}{\partial v}
+ \frac{C^{2}_{1}}{4 \pi} \frac{\partial t}{\partial v} = 0, \\
&&\frac{\partial z}{\partial v} - v \frac{\partial t}{\partial v} +
\rho \frac{\partial t}{\partial \rho} = 0. \label{lineq}
\end{eqnarray}
To simplify the notation we drop the index of $v_z$. Eliminating the derivatives
of $z$ we obtain one equation for $t(\rho, v)$
\begin{equation}
\rho \frac{\partial^2 t}{\partial \rho^2} + 2 \frac{\partial t}{\partial \rho}
 - C_{2} \Gamma \rho^{\Gamma - 2} \frac{\partial^2 t}{\partial v^2} -
\frac{C^{2}_{1}}{4 \pi} \frac{\partial^2 t}{\partial v^2} = 0. \label{oneeq}
\end{equation}
Solution of this equation cannot be expressed in terms of known special
functions for the arbitrary values of $\Gamma \ne 2$. However, we can obtain
a quasiclassical approximation to the general solution after applying Laplace
transform to the variable $v$.
When $v>0$, the quasiclassical solution to~(\ref{oneeq}) is
\begin{equation}
\label{quasit}
t(\rho, v) = -\frac{1}{2 \rho \sqrt{\xi(\rho)}} \int_{\rho_{-}}^{\rho_{+}}
\frac{f(\rho_1)}{\sqrt{\xi(\rho_1)}} d\rho_1.
\end{equation}
Here $\xi(\rho) = \left(C_2 \Gamma \rho^{\Gamma - 3} + C_1^2/(4
\pi \rho)\right)^{1/2}=\left(\Gamma P + B^2/(4 \pi)\right)^{1/2}
\rho^{-3/2} = c_s/\rho,$ and $c_s$ is the fast magnetic sound speed.
The function $f(\rho)$ appearing in the solution depends on the initial
conditions. We may assume that the velocity $v$ is zero when $t=0$.
Accordingly, $\partial v/\partial z|_{t=0}=0$ as well. It is convenient to
write the initial condition for the density $\rho$ at $t=0$ in the following
way $\int_{\rho_0}^{\rho} f(\rho) d\rho = z$, where $\rho_0$ is the density
at $z=0$. From the continuity equation of system~(\ref{twoeq}) it follows that
$\partial \rho/\partial t|_{t=0} = 0$. From the first equation of the same
system (the equation of motion) we see that $\partial v/\partial t|_{t=0} =
-\rho \xi^2(\rho)/f(\rho)$.
After we change variables to $(t(v, \rho)$, $z(v, \rho))$, our initial
conditions become
\begin{eqnarray*}
&&\left.\frac{\partial t}{\partial v}\right|_{v=0} = - \frac{f(\rho)}{\rho
\xi^2(\rho)};\quad
\left.\frac{\partial t}{\partial \rho}\right|_{v=0} = 0;\\
&&\left.\frac{\partial z}{\partial v}\right|_{v=0} = 0;\quad
\left.\frac{\partial z}{\partial \rho}\right|_{v=0} = f(\rho);\quad
\left.t\right|_{v=0} = 0.
\end{eqnarray*}
In the obtained solution,
$\rho_+$ satisfies the equation $v = \int_\rho^{\rho_+} \xi(\rho_1) d\rho_1$,
and $\rho_-$ satisfies the equation $v = \int_{\rho_-}^\rho \xi(\rho_1)
d\rho_1$. For $v > \int_0^\rho \xi(\rho_1) d\rho_1$ the value $\rho_-$
cannot be determined, because at the limiting value $v = \int_0^\rho
\xi(\rho_1) d\rho_1$ the density $\rho_-$ vanishes and cannot be negative
later. Since $\xi=c_s/\rho$, the condition $v > \int_0^\rho \xi(\rho_1)
d\rho_1$ is virtually identical to $v > c_s$, and quasiclassical
solution~(\ref{quasit}) is not valid.

Substituting~(\ref{quasit}) into~(\ref{lineq}) gives the following
approximation for $z(v,\rho)$
\begin{equation}
\label{quasiz}
z(v,\rho)=vt(v,\rho) - \frac{\partial}{\partial \rho}\left[\rho \int_0^v
t(v_1, \rho) dv_1\right] + \int_{\rho_0}^\rho f(\rho_1) d\rho_1.
\end{equation}
The integrals appearing in expressions~(\ref{quasit},\ref{quasiz}) have their
simplest form when $\Gamma=2$, which is close to the adiabatic exponent
for an ideal gas $\Gamma=5/3$. In this case the solution is
\begin{eqnarray}
&&t(v, \rho) = - \frac{4 f_0}{\alpha^2} v \rho^k, \nonumber \\
&&z(v, \rho) = \frac{f_0 (\rho^{k+1}-\rho_0^{k+1})}{k+1} + \frac{(k-1) \alpha^2}
{8 f_0} \rho^{-k} t^2. \label{quasi2}
\end{eqnarray}
Here we have chosen the function $f(\rho)$ to be the power function
$f_0 \rho^k$ where $k$ is an arbitrary value and $\alpha = 2(2 C_2 + C_1^2/4 \pi)^{1/2}$.
Solutions $t(v,\rho)$ and $z(v,\rho)$~(\ref{quasi2}) are always single-valued
functions of ($v,\rho$), but the functions $v(t,z)$ and $\rho(t,z)$ may not be.
The quantity $\rho(t,z)$ can be found by reversing the second equation of
system~(\ref{quasi2}). This can be uniquely done until the moment when the shock occurs
and the Jacobian of the transformation of the unknown variables vanishes
\[
J = \frac{\partial(t, z)}{\partial(\rho, v)} = \frac{4 f_0^2}{\alpha^2}
\rho^{2 k} \left( 1 + 2k(1-k)\frac{v^2}{\alpha^2 \rho} \right) = 0.
\]
For $k<0$ and $k>1$ the Jacobian $J$ always vanishes at a value
$q=v^2/(\alpha^2 \rho)$. When the shock occurs, the derivatives $\partial
\rho/\partial z$ and $\partial v/\partial z$ become infinite which leads to the
singularity in the electromagnetic field. The question whether
there is a shock for almost arbitrary initial perturbations
is very important, so we would like to obtain an exact solution of
the MHD equations, and it is only possible for $\Gamma=2$.
This solution can be expressed in terms of hypergeometric functions
\begin{eqnarray}
&&t(\rho, v) = - \frac{4 f_0}{\alpha^2} v\rho^k {_2F_1}
\left(-1-k, -k, \frac{3}{2}, \frac{v^2}{\alpha^2 \rho}\right), \nonumber\\
&&z(\rho, v) = \frac{f_0 \rho^{k+1}}{k+1} {_2F_1}\left(-2-k, -1-k,
\frac{1}{2}, \frac{v^2}{\alpha^2 \rho}\right) - \label{exact-tz} \\
&&\frac{6 f_0 \rho^k v^2}{\alpha^2} {_2F_1}\left(-1-k, -k, \frac{3}{2},
\frac{v^2}{\alpha^2 \rho}\right)
-\frac{f_0 \rho_0^{k+1}}{k+1}. \nonumber
\end{eqnarray}
The hypergeometric function $_2 F_1(a,b,c,q)$ appearing in these equations
has a branch point at $q=1$, which means
that solution~(\ref{exact-tz}) cannot be continued into the region $q>1$.

Note that quasiclassical solution~(\ref{quasi2}) for $\Gamma=2$ corresponds
to the first two terms in the expansion of exact solution~(\ref{exact-tz}) into powers
of $q$.

Both in the exact solution and in the quasiclassical one, the singularity
occurs for $0<q<1$ when the Jacobian $J$ vanishes (the shock point).
However, in the exact solution the shock occurs only for negative values
of $k$, namely for $k < -1/2$. These values of $k$ correspond to the more
natural initial conditions when the initial value of the acceleration of
the sheet $\partial v/\partial t|_{t=0} \propto \rho^{-k}$ does not have
a singularity at $\rho=0$. The behavior of the Jacobian $J$ of
$q=v^2/(\alpha^2 \rho)$ for the exact and quasiclassical solutions is
shown in Fig.2.

\begin{figure*}
\centering
\includegraphics[width=84mm]{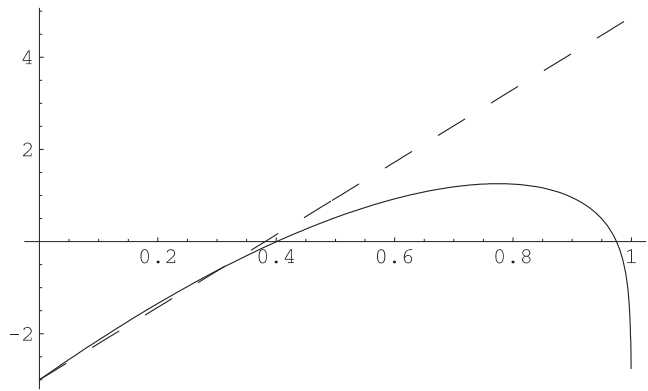}
\caption{$- 3 \alpha^2 J/4 f_0^2 \rho^{2k}$ of $q$ at $k=-3/4$,
the dashed line is the quasiclassical approximation}
\end{figure*}

The figure shows that the approximate solution describes the behavior of the
sheet well right up to the shock point, so later we will use only
quasiclassical solution~(\ref{quasi2}).

An expression for the $y$ component of the plasma velocity should be
added to formulas~(\ref{quasi2}). It can be easily found from
the equation $\partial v_{y}/\partial t + v \partial v_{y}/
\partial z = 0$, whose
solution is an arbitrary function of the variable $\rho^{k+2} \eta$,
$v_y = \Phi(\rho^{k+2} \eta)$, where
\begin{equation}
\label{etadef}
\eta = 1 + \frac{k (k+2) \alpha^2}{8 f_0^2} \rho^{-2k-1} t^2 = 1 + 2k(k+2)q.
\end{equation}
The function $\Phi$ is determined by an initial condition for
the motion of the sheet in the $y$ direction at $t=0$. The initial velocity $v_y$ can
be arbitrary for an infinitely wide sheet. But since the real
geometry of our problem is not flat, we may assume that
$v_y \sim v_z = v$.

The electric and magnetic fields are given by
\begin{equation}
\label{fields}
B_x \equiv B = C_1 \rho,\,
\frac{E_y}{B} = -\frac{v}{c} = \frac{\alpha^2}{4 f_0 c} \rho^{-k} t,\,
\frac{E_z}{B} = \frac{v_y}{c} = \frac{\Phi}{c},
\end{equation}
and the other components vanish.

\section[]{Acceleration of charged particles}

Evolution of the current sheet causes a drastic change of the magnetic field,
its energy, and the appearance of the electric field.
Due to the high speed of this process (it lasts for only about $L/c_s\simeq 0.1 sec$),
the energy of the magnetic field is passed to the energy of the accelerated particles.
Acceleration is most effective in a region with the fastest evolution,
i.e. near the singularity.
Let us consider acceleration of the charged particles in the vicinity of
singularity. We assume that these particles are relativistic, so that
$\varepsilon = cp$, and that their fraction is small compared to the plasma
density in the sheet. The singularity appears in the above solution of the
MHD equations during the compression of the current sheet.
The distribution function $F_*(t,z,p_x,p_y,p_z)$ of relativistic
particles in Cartesian coordinates satisfies collisionless Lioville's
equation
\[
\frac{\partial F_*}{\partial t} + \frac{c^2}{\varepsilon} p_z
\frac{\partial F_*}{\partial z} + {\cal F}_i \frac{\partial F_*}
{\partial p_i} = 0.
\]
Let us change the Cartesian coordinates $(p_x, p_y, p_z)$ to the cylindrical
ones $(p_\parallel, p_\perp, \varphi)$ in the momentum space.
Here $p_\parallel$ is a component of the particle momentum along the magnetic
field $B=B_x$, $p_\perp$ is that perpendicular to the magnetic field, and
$\varphi$ is a rotation angle of the particle around the magnetic field,
\[
p_x = p_\parallel,\,
p_y = p_\perp \sin{\varphi},\,
p_z = p_\perp \cos{\varphi}.
\]
The force acting on the particles is the Lorentz force
\[
{\bf {\cal F}} = e \left({\bf E} + \frac{c}{\varepsilon}
[{\bf p} {\bf B}] \right).
\]
Thus we have the kinetic equation
\begin{eqnarray*}
&&\frac{\partial F_*}{\partial t} +\frac{c p_\perp}{p} \cos{\varphi}
\frac{\partial F_*}{\partial z} + e \left(E_y \sin{\varphi} + E_z
\cos{\varphi} \right) \frac{\partial F_*}{\partial p_\perp} +\\
&&\left[ \frac{e}{p_\perp} \left(E_y \cos{\varphi} - E_z \sin{\varphi}
\right) + \omega_c \right] \frac{\partial F_*}{\partial \varphi} = 0,
\end{eqnarray*}
where $\omega_c$ is the cyclotron frequency of the particle rotation,
$\omega_c = eB/p$.
Let us expand the function $F_*$ of $\varphi$ into a Fourier series and
leave only the first 3 terms
\[
F_* = F_0 + F_c \cos{\varphi} + F_s \sin{\varphi}.
\]
Hence we get the system of equations for the functions $F_0$, $F_c$, $F_s$
\begin{eqnarray*}
&&\frac{\partial F_{0}}{\partial t} + \frac{c p_\perp}{2 p} \frac{\partial
F_{c}}{\partial z} + \frac{1}{2} \frac{e E_y}{p_\perp} F_{s} +
\frac{e E_y}{2} \frac{\partial F_{s}}{\partial p_\perp} +\\
&&\frac{1}{2} \frac{e E_z}{p_\perp} F_{c} + \frac{e E_z}{2} \frac{\partial F_{c}}
{\partial p_\perp} = 0, \\
&&\frac{\partial F_{c}}{\partial t} + \frac{c p_\perp}{p} \frac{\partial F_{0}}
{\partial z} + e E_z \frac{\partial F_{0}}{\partial p_\perp} + \omega_c F_{s}
= 0, \\
&&\frac{\partial F_{s}}{\partial t} + e E_y \frac{\partial F_{0}}{\partial
p_\perp} - \omega_c F_{c} = 0.
\end{eqnarray*}
We need only know the function $F_0(t,z,p_\parallel,p_\perp)$,
which is independent of $\varphi$, because the other components of the
distribution function make no contribution after the averaging over
the rotation angle $\varphi$.
We solve the second and the third equations for $F_c$, $F_s$, and substitute
the result into the first equation.
This procedure is simplified in two cases: (a)
$\omega_c F_i \gg \partial F_j/\partial t$, (b) $\omega_c F_i \ll \partial F_j/\partial t$.

In the first case, the cyclotron period is much less than the time scale of
evolution of the distribution function. This condition is only violated close to
the center of the sheet, where the magnetic field is zero and particles
are not magnetized. The magnetic field is very strong in the case of
the interaction between the supernova shock and the magnetized star
($B\sim 10^6$ Gauss), therefore the region, where the particles are not
magnetized, is negligibly small. Hence we may assume that the condition (a) holds everywhere.
Substituting
\[
F_s = -\frac{E_z}{B} p \frac{\partial F_0}{\partial p_\perp} -
\frac{cp_\perp}{eB} \frac{\partial F_0}{\partial z},
F_c = \frac{E_y}{B} p \frac{\partial F_0}{\partial p_\perp}
\]
into the first equation, we obtain
\begin{equation}
\label{kineq1}
\frac{\partial F_{0}}{\partial t} - c \frac{E_y}{B} \frac{\partial F_{0}}
{\partial z} + \frac{cp_{\perp}}{2} \frac{\partial}{\partial z} \left(
\frac{E_y}{B}\right)\frac{\partial F_{0}}{\partial p_{\perp}} = 0.
\end{equation}
Equation~(\ref{kineq1}) has a simple physical meaning. All particles
move in the $z$ direction with a velocity of the electric drift $-c E_y/B$,
moreover the particles have an acceleration in this direction due to the
variation of this velocity along the $z$ axis, and hence gain the transverse
momentum $p_\perp$. It is important to note that the acceleration
$\partial(c E_y/B)/\partial z$ tends to infinity at the shock point,
\[\frac{\partial}{\partial z} \left(\frac{E_y}{B}\right) = - \frac{4kq}{t}
\left(1 + 2k(1-k)q\right)^{-1}.
\]
The dependence of $E_y/B$ on the coordinate $z$ and the time $t$ is given
by the formula~(\ref{fields}).
Now, it will be convenient to change variables from $(t,z,p_\perp)$ to $(\eta,
\rho,p_\perp)$. In the new variables~(\ref{kineq1}) reads
\begin{equation}
\label{kineq2}
(k+2) \eta \frac{\partial F_0}{\partial \eta}
- \rho \frac{\partial F_0}{\partial \rho} - \frac{1}{4} p_\perp \frac
{\partial F_0}{\partial p_\perp} = 0.
\end{equation}
We may assume that the distribution function of fast plasma particles
is Maxwellian at $t=0$ ($\eta=1$), and their density is
proportional to the plasma density, i.e. the fraction of the accelerated
injected particles is constant along the entire sheet
\[
F_0(\eta=1,\rho,p_\perp)=c_0 \rho e^{-c p_\perp/T_0}.
\]
The solution of kinetic equation~(\ref{kineq2}) with such initial conditions is
\[
F_0(\eta,\rho,p_\perp)=c_0 \rho \eta^{1/(k+2)} \exp{\left(-c p_\perp
\eta^{1/(2k+4)}/T_0 \right)}.
\]
It shows that the distribution function of the particles
remains Maxwellian, but the temperature is time-dependent $T=T_0
\eta^{-1/(2k+4)}$. For $-2<k<0$ the temperature is rising as $\eta$ tends to
zero. At the moment $\eta=0$ the temperature becomes infinite, which means
that the distribution function is constant in the region $p_\perp<\infty$.
Note that the moment $\eta=0$ occurs before the shock for $-1/2<k<0$ and
these moments coincide if $k=-1/2$. The case $k=-1/2$ corresponds to the
initial distribution of density $\rho(z)$ with constant initial plasma
temperature along the entire sheet. Though formally the particles
are accelerated up to infinite energies, it is clear that the cyclotron
radius of the fastest particles should be less than the typical thickness of
the sheet $L$, and also, that their cyclotron frequency should be greater than
the inverse typical evolution time $\Delta t$ of the sheet. Therefore, we
have two estimates for $p_{\perp m}$: $p_{\perp m} = p_{1 m}=eL<B>/c$
and $cp_{\perp m}=cp_{2 m} = T_0\Delta\eta^{-1/(2k+4)}$,
where $\Delta\eta$ can be found from the definition of $\eta$ and is
estimated to be $\Delta\eta\simeq (c_s/(L\omega_{c0}))^2(p_{2 m}/mc)^2$,
\quad $\omega_{c0}=eB/mc$. Finally we have $p_{2 m}/mc =
(T_0/(mc^2))^{(k+2)/(k+3)}(L\omega_{c0}/c_s)^{1/(k+3)}$. Later we will see
that $p_{1 m}>>p_{2 m}$ for the actual values of the parameters,
so we can assume that $p_{m} = p_{2 m}$.

\section[]{Synchrotron radiation}

The most interesting case for us is when $k=-1/2$, and in this case all plasma
masses up at the moment of shock near the two planes $z_{+,-} = \pm 2 f_0
\sqrt{\rho_0}$, symmetric with respect to the center of the sheet.
The estimate for $z_+$ is $2 c_{s0}^2/|\partial v/\partial t||_{t=0}$, it is
reciprocal to the initial acceleration of the sheet
$\partial v/\partial t|_{t=0}$, and, of course, $z_+$ should be less than
the thickness of the sheet $L$. The plasma velocity $v_z$ is positive
for $z > 0$ and $v_z < 0$ for $z < 0$, i.e. plasma particles move away from
the center of the sheet into the regions around $z_{+,-}$ with a strong
magnetic field. Due to the strong magnetic field, the synchrotron radiation
becomes significant for the plasma particles (electrons) there.
For $z > z_+$ and $z < z_-$ the quasiclassical approximation used in the
solution of the MHD equations is not applicable. We assume that the magnetic
field is almost constant at the large distances $z > z_+$ and $z < z_-$ from
the center of the sheet, and its value is equal to the magnetic field at the
boundary of the sheet. The electric field with the components $E_y$ and
$E_z$ exists there as well, and plasma velocity in the electromagnetic
field should be $v \sim c_s$, where $c_s$ is the fast magnetic
sound speed.

Thus, fast particles with the constant distribution function
obtained above move into the regions with constant fields due to the electric
drift and strongly radiate there. Next, we find $F$, the distribution function
of these particles, and their radiation intensity $I$.

We may assume that $E_y/B \ll 1$ and $E_z/B \ll 1$, i.e. the plasma motion is
non-relativistic. In this case the drift velocity is
${\bf v}_{dr} = (0, cE_z/B, -cE_y/B)$.

The function $F$ satisfies the following continuity equation in the phase space
\[
\frac{\partial F}{\partial t} + c \frac{p_z}{p} \frac{\partial F}{\partial z}
+ \frac{\partial {\cal F}_i}{\partial p_i} F +  {\cal F}_i \frac{\partial F}
{\partial p_i} = 0.
\]
The force ${\cal F}_i$ acting on the particles has the simplest expression
in the reference frame moving with the velocity of the electric drift, where
there is no electric field. Since the drift is non-relativistic, this force
is the same in the stationary reference frame.
We use tildes for the values observed in the moving frame.
The friction force due to radiation is~\cite{LL21}
\[
{\bf F}_{fr} = - \frac{2 e^4 \tilde{B}^2}{3 m^4 c^6} \frac{1}
{\tilde{p}} \left( \tilde{p}_y^2 + \tilde{p}_z^2 \right)
\left(\tilde{p}_x,\tilde{p}_y,\tilde{p}_z\right),
\]
note that the magnetic field in the moving reference frame is $\tilde{B} =
B \left( 1 - (E_y^2+E_z^2)/B^2 \right)$,
and components of the electric field are $\tilde{E}_y \equiv 0$,
$\tilde{E}_z \equiv 0$.
Let us denote
\[
\kappa = \frac{2 e^4}{3 m^4 c^6}.
\]
The net force acting on the particles is
\begin{eqnarray*}
&&{\cal F}_x = {\cal F}_{fr,x} ,\\
&&{\cal F}_y = e \tilde{B} \frac{\tilde{p}_z}{\tilde{p}} +
{\cal F}_{fr,y}, \\
&&{\cal F}_z = -e \tilde{B} \frac{\tilde{p}_y}{\tilde{p}} + {\cal F}_{fr,z}.
\end{eqnarray*}
Again, change the Cartesian coordinates $(p_x, p_y, p_z)$ to the cylindrical
ones
$(p_\parallel, p_\perp, \varphi)$ in the momentum space
\begin{eqnarray*}
&&p_x = \tilde{p}_x = p_\parallel \\
&&p_y = \frac{E_z}{B} \tilde{p} + \tilde{p}_y = \frac{E_z}{B} \tilde{p} +
p_\perp \sin{\varphi} \\
&&p_z = -\frac{E_y}{B} \tilde{p} + \tilde{p}_z = -\frac{E_y}{B} \tilde{p} +
p_\perp \cos{\varphi}.
\end{eqnarray*}
Notice that a non-relativistic momentum addition rule was used here
as the drift is non-relativistic.

Further, we neglect terms of the order higher than $E_y/B$ and $E_z/B$ in all
calculations. Under this approximation $\tilde{B}=B$,
\begin{equation}
{\cal F}_i \frac{\partial F}{\partial p_i} = {\cal F}_i \frac{\partial F}
{\partial \tilde{p}_i} - \frac{{\cal F}_y \tilde{p}_y E_z + {\cal F}_z
\tilde{p}_z E_y}{B \tilde{p}} \frac{\partial F}{\partial \tilde{p}_y} +
\frac{{\cal F}_y \tilde{p}_y E_y + {\cal F}_z \tilde{p}_z E_y}{B \tilde{p}}
\frac{\partial F}{\partial \tilde{p}_z}.
\end{equation}
Since divergence is a scalar value and the Lorentz force is solenoidal, we
have
\[
\frac{\partial {\cal F}_i}{\partial p_i} = \frac{\partial {\cal F}_i}
{\partial \tilde{p}_i} = \frac{\partial {\cal F}_{fr,i}}{\partial
\tilde{p}_i} = -4 \kappa B^2 \frac{p_\perp^2}{\tilde{p}}.
\]
Substituting the expressions obtained above in the original kinetic equation,
and averaging it over the rotation angle $\varphi$, we obtain
\begin{equation}\label{keq}
\frac{\partial F}{\partial t} -\frac{c E_y}{B} \frac{\partial F}
{\partial z} - \kappa B^2 \frac{p_\perp^2}{p}\left( p_{\parallel}
\frac{\partial F}{\partial p_{\parallel}} + p_{\perp} \frac{\partial F}
{\partial p_{\perp}} \right) -
4\kappa B^2 \frac{p_{\perp}^2}{p} F = 0.
\end{equation}
The identity $\tilde{p}=p$ is used here, it can be proved by expanding $\tilde{p}$
into a Taylor series in powers of $E_y/B$ and $E_z/B$ and
averaging over the angle $\varphi$.

Let us consider the case $z > 0$ ($z < 0$ is analogous due to the symmetry of
the problem).
The particles with constant initial distribution function move into the region $z>z_+$
after accumulating at $z_+ = 2 |f_0| \sqrt{\rho_0}$, according to the kinetic equation
obtained above. We can suppose without loss of
generality that the origin of the $z$ coordinate is at the point $z_+$, i.e.
we can solve the kinetic equation with the initial condition
\[
F(t=t_*; z=0; p_\parallel; p_\perp) = c_1 \Theta(p_m -p).
\]
Here $t_*$ is the time of the constant distribution formation.
Let us notice that that the constant distribution function of
accelerated particles obtained in the previous
section was the function of the transverse momentum $p_\perp$. Here, in
contrast, we use the isotropic function with the same parameters. The reason
is that the plasma with strongly anisotropic particle distribution is very
unstable with respect to the excitation of the electrostatic waves leading
to fast isotropization. The growth rate of such instabilities is of order
of the cyclotron frequency $\omega_c$ or the plasma frequency
$\omega_p$~\cite{BIP}.
We are looking for a stationary distribution function (the solution of~(\ref{keq})
which does not depend on time) to find the spectrum of the radiation of the particles.
This solution is
\[
F(z; p_\parallel; p_\perp) = \frac{c_1}{\left( 1 + \frac{\kappa B^3 z
p_\perp^2}{ c E_y p} \right)^4}\Theta\left( p_m - \frac{p}{1 +
\frac{\kappa B^3 z p_\perp^2}{ c E_y p}} \right).
\]
Having found the distribution function, we can compute the spectrum and the
directivity of the radiation, i.e. dependence of the radiation intensity on
the direction and on the frequency.

First, let us find the synchrotron radiation from one particle moving with
the velocity $\bf{v}$. The prime denotes that the variable is observed
in the coordinate frame $K'$ moving with the velocity $v_\parallel$.
To find the radiation, we will need to change the coordinate frame
from the stationary one to $K'$ and back. We denote the angle between the
direction of the magnetic field and the velocity of the particle by
$\chi$ and the angle between the plane of the particle rotation and the
direction of the radiation by $\theta$. The radiation in the $K'$ coordinate
frame is
$$
dI^\prime = d\Omega^\prime \frac{e^4 {B^\prime}^2 v_\perp^{\prime 2}
(1 - v_\perp^{\prime 2}/c^2)}{8 \pi^2 m^2 c^5} \times
\frac{\left[ 2 - \cos^2{\theta^\prime} - \frac{v_\perp^{\prime 2}}{4 c^2}
\left(1+\frac{3 v_\perp^{\prime 2}}{c^2}\right)
\cos^4{\theta^\prime} \right]}{\left( 1 - \frac{v_\perp^{\prime 2}}{c^2}
\cos^2{\theta^\prime} \right)^{7/2}}.
$$
Here
\[
d\Omega^\prime = 2 \pi \cos{\theta^\prime} d\theta^\prime,\, B^{\prime}=
\gamma B,\, v_\perp^\prime = \gamma v_\perp;
\]
\[
v_{\perp} = v \sin{\chi},\, v_{\parallel} = v \cos{\chi},\,
\gamma = 1/\sqrt{1-\beta^2},\, \beta = v_{\parallel}/c.
\]
Therefore, the radiation in the stationary coordinate frame is
\begin{eqnarray}
&&dI = (1 + \beta \sin{\theta^\prime}) dI^\prime,\nonumber\\
&&\sin{\theta} = \frac{\beta + \sin{\theta^\prime}}{1 +
\beta \sin{\theta^\prime}},
\label{angletransform}\\
&&\frac{d\theta}{\cos^2{\theta}} = \gamma (1 + \beta \sin{\theta^\prime})
\frac{d\theta^\prime}{\cos^2{\theta^\prime}},\nonumber
\end{eqnarray}
$$
\frac{dI}{d\theta} = \gamma^3 \frac{\cos^3{\theta^\prime}}{\cos^2{\theta}}
\frac{e^4 B^2 v_\perp^2 (1 - \gamma^2 v_\perp^2/c^2)}{4 \pi m^2 c^5} \times
\frac{\left[ 2 - \cos^2{\theta^\prime} - \frac{\gamma^2 v_\perp^2}{4 c^2}
\left(1+\frac{3 \gamma^2 v_\perp^2}{c^2}\right)
\cos^4{\theta^\prime} \right]}{
\left( 1 - \gamma^2 \frac{v_\perp^2}{c^2} \cos^2{\theta^\prime} \right)^{7/2}}.
$$
The sought dependence of the radiation intensity on the direction
$\theta$ is
\begin{eqnarray}
&&\langle \frac{dI}{d\theta} \rangle = \int_0^{z_{max}} dz \int_{p_{min}}^{
\infty} p^2 dp \int_0^{\pi} \gamma^3 \frac{\cos^3{\theta^\prime}}{\cos^2 {\theta}}\times
\frac{e^4 B^2 v_\perp^2 (1 - \gamma^2 v_\perp^2/c^2)}{4 \pi m^2 c^5} \times\nonumber\\
&&\frac{\left[ 2 - \cos^2{\theta^\prime} - \frac{\gamma^2 v_\perp^2}{4 c^2}
\left(1+\frac{3 \gamma^2 v_\perp^2}{c^2}\right) \cos^4{\theta^\prime} \right]}
{\left( 1 - \gamma^2 \frac{v_\perp^2}{c^2} \cos^2{\theta^\prime} \right)^{7/2}}
\times\nonumber\\
&&\frac{c_1}{\left(1+\frac{\kappa B^3 z}{c E_y} p \sin^2{\chi}\right)^4}
\Theta\left( p_m - \frac{p}{1+\frac{\kappa B^3 z}{c E_y} p \sin^2{\chi}}
\right)\times 2 \pi \sin{\chi} d\chi.
\label{direction}
\end{eqnarray}
Note, that only the particles with $\gamma \gg 1$ make significant contribution to the
radiation. In the $K'$ coordinate frame the synchrotron radiation is
concentrated near the plane of the particle rotation, which means that
$\theta^\prime=0$. The most part of the radiation is concentrated within
the angle $\Delta\theta^\prime \simeq 1/\gamma \ll 1$. Note that
$\Gamma=(1-v^2/c^2)^{-1/2} \gg 1$ since $\gamma \gg 1.$

From Equation~(\ref{angletransform}) it follows that $\sin{\theta}
\sim \beta = \frac{v}{c} \cos{\chi} \simeq \cos{\chi}$.
Hence the particles with the pitch angle $\chi$ radiate in the stationary
coordinate frame mainly into the angle
\begin{equation}
\theta = \frac{\pi}{2} - \chi.
\label{theta-chi}
\end{equation}
Let us perform the Taylor expansion of all expressions appearing in the
integral over $\chi$ in~(\ref{direction}) in powers of $1 - v^2/c^2$ and
leave only the first, most essential terms. Using the smallness of the angle
$\theta^\prime$, we get
\[
\gamma \simeq 1/\sin{\chi},\, \cos{\theta^\prime} \simeq 1,\,
v_\perp \simeq c \sin{\chi},
\]
\[
(1 - \gamma^2 v_\perp^2/c^2) \simeq \left( 1 - \gamma^2 \frac{v_\perp^2}
{c^2} \cos^2{\theta^\prime} \right) \simeq \frac{1}{\Gamma^2 \sin^2{\chi}},
\]
\[
\left[ 2 - \cos^2{\theta^\prime} - \frac{\gamma^2 v_\perp^2}{4 c^2}
\left(1+\frac{3 \gamma^2 v_\perp^2}{c^2}\right) \\
\cos^4{\theta^\prime} \right] \simeq \frac{7}{4 \Gamma^2 \sin^2{\chi}}.
\]
Substituting these equations into the integral~(\ref{direction}) and
using~(\ref{theta-chi}), we obtain that the integrand is approximately equal to
$$
\frac{7}{8} c_1 \frac{e^4 B^2}{m^2 c^3} \Gamma^3 \frac{\cos{\theta}}
{\left(1+\frac{\kappa B^3 z}{c E_y} p \cos^2{\theta}\right)^4} \times
\Theta\left( p_m - \frac{p}{1+\frac{\kappa B^3 z}{c E_y} p \cos^2{\theta}}
\right) d\chi.
$$
Numerical computation shows that the radiation at the angle $\theta$ is
produced mainly by the particles with $\Delta \chi \simeq 1.52 / \Gamma$,
i.e. the integral over $d\chi$ is close to
$$
1.33 c_1 \frac{e^4 B^2}{m^2 c^3} \Gamma^2 \frac{\cos{\theta}}{\left(1+\frac
{\kappa B^3 z}{c E_y} p \cos^2{\theta}\right)^4} \times
\Theta\left( p_m - \frac{p}{1+\frac{\kappa B^3 z}{c E_y} p \cos^2{\theta}} \right).
$$
The upper limit of the integration over $p$ in Equation~(\ref{direction}) is
\[
p = p_{cr} = \frac{p_m}{1 - \frac{\kappa B^3 z}{c E_y} p_m \cos^2{\theta}}.
\]
Notice that $p_{cr} < p_m$ since $E_y/B < 0$. The lower limit of the
integration, $p_{min}$, can be replaced by zero. Hence
$$
\langle \frac{dI}{d\theta} \rangle =1.33 c_1 \frac{e^4 B^2}{m^4 c^5}
\cos{\theta} \int_0^{z_{max}} dz \int_0^{p_{cr}}p^4 \times
\left(1+\frac{\kappa B^3 z} {c E_y} p \cos^2{\theta}\right)^{-4} dp.
$$
These two integrals can be evaluated analytically:
\begin{eqnarray*}
&&\langle \frac{dI}{d\theta} \rangle = 1.33 c_1 \frac{e^4 B^2}{12 m^4 c^5}
\left( \frac{p_m}{K} \right)^5 z_{max}\frac{1}{\cos{\theta}} \left( 3K^4 -\right.\\
&&\left.\frac{4K^3}{\cos^2{\theta}} + \frac{6K^2}{\cos^4{\theta}} - \frac{12K}
{\cos^6{\theta}} + \frac{12}{\cos^8{\theta}} \log(1+K \cos^2{\theta}) \right),
\end{eqnarray*}
where $K = - (2 e^4 B^3/3 m^4 c^7 E_y) p_m z_{max} \gg 1$. Angular
distribution of the intensity $\langle dI/d\theta \rangle$ is maximal at
$\cos{\theta} \simeq K^{-1/2}$, it vanishes at $\cos{\theta}=0$, and has the
width maximum $\Delta(\cos{\theta})$ close to $K^{-1/2}$.
That is, the radiation is strongly collimated along the magnetic field and
the collimation angle $\Delta \theta$ can be estimated as $K^{-1/2} \ll 1$.
Let us find the radiation spectrum. For a single particle the spectral
intensity of the synchrotron radiation is~\cite{LL22}
\[
\frac{dI}{d\omega} = \frac{\sqrt{3}}{2 \pi} \frac{e^3 B_\perp}{mc^2}
F\left( \frac{\omega}{\omega_c} \right).
\]
Here
\[
F(\xi) = \xi \int_{\xi}^{\infty} K_{5/3}(\xi) d\xi,
\]
\[
\omega_c = \frac{3 e B_\perp}{2mc} \left( \frac{\varepsilon}{mc^2}
\right)^2 \simeq \frac{3eB}{2 m^3 c^3} p^2 \sin{\chi},
\]
\[
B_\perp = B \sin{\chi}.
\]
The total spectrum is given by
\begin{equation}\label{spectrum}
\langle \frac{dI}{d\omega} \rangle = c_1\int^\pi_0 2 \pi \sin{\chi}
d\chi \int^{z_{max}}_0 dz \times
\int^{p_{cr}}_0 \left(1+\frac{\kappa B^3 z}{c E_y}
p \sin^2{\chi}\right)^{-4}\frac{dI}{d\omega} p^2 dp,
\end{equation}
where
\[
p_{cr} = p_m\left(1 - \frac{\kappa B^3 z}{c E_y} p_m \sin^2{\chi}\right)^{-1}.
\]
Asymptotic behavior of the function $F(\xi)$ is known to be
\[
F(\xi) \sim \sqrt{\frac{\pi \xi}{2}} e^{-\xi} \,\mbox{as} \, \xi \to \infty.
\]
Using Laplace's method, we have the asymptotics when $\omega \to \infty$
\begin{eqnarray*}
&&\int^{p_{cr}}_{p_{min}}\left(1+\frac{\kappa B^3 z}{c E_y} p \sin^2{\chi}
\right)^{-4} \frac{dI}{d\omega} p^2 dp \simeq \frac{3}{8} \times\\
&&\left(\frac{e^7 B^3}
{\pi m^5 c^7 \omega}\right)^{1/2}p_m^4 \sin^{3/2}{\chi}\exp{\left( -\omega
\frac{2 m^3 c^3}{3 eB} \frac{1}{p_{cr}^2 \sin{\chi}} \right)}.
\end{eqnarray*}
After substituting the expression for $p_{cr}$, we obtain the asymptotics for the two other
integrals when $\omega \to \infty$
\begin{eqnarray*}
&&\int^{z_{max}}_0 dz \int^{p_{cr}}_{p_{min}} (...) dp \sim
\frac{9c_1}
{32\pi^{1/2}} \omega^{-3/2} p_m^3 \times\\
&&\left(\frac{m^3 c^7 E_y^2}{e B^3
\sin{\chi}}\right)^{1/2} \exp{\left( -\omega \frac{2 m^3 c^3}{3 eB}
\frac{1}{p_m^2 \sin{\chi}} \right)}.
\end{eqnarray*}
Finally, we obtain the asymptotic of the spectral intensity
\[
\langle \frac{dI}{d\omega} \rangle \simeq \frac{27 \pi \sqrt{3}}{32}
c_1 p_m^6 \frac{e |E_y|}{m^3 c} \frac{1}{\omega^2} \exp{\left( -\omega
\frac{2 m^3 c^3}{3eB p_m^2} \right)} \, \mbox{as} \, \omega \to \infty.
\]
The exponential damping of the spectrum begins at the frequency
\[
\omega_* = \frac{3}{2} \frac{e B}{mc} \left(\frac{p_m}{mc}\right)^2.
\]
Now let us find the asymptotics of the spectrum as $\omega \to 0$.
Recall that the radiation is mainly concentrated in the angle $K^{-1/2}$
radian. Only the particles with $\chi \sim K^{-1/2}$ make a substantial
contribution (equation~(\ref{theta-chi})). The asymptotics of $F(\xi)$ is
\[
F(\xi) \sim 2^{2/3} \Gamma\left(\frac{2}{3}\right) \xi^{1/3} \, \mbox{as}
\,\xi \to 0.
\]
Therefore, when $\omega \to 0$, we have
$$
\int^{p_{cr}}_{p_{min}} \frac{dI}{d\omega}p^2\left(1+\frac{\kappa B^3 z}
{c E_y} p \sin^2{\chi}\right)^{-4} dp \simeq
\frac{3^{7/6} \Gamma(2/3)}{7 \pi}\frac{e^{8/3} B^{2/3}}{c} \omega^{1/3} p_{cr}^{7/3} \sin^{2/3}{\chi}.
$$
Finally, after integrating over $\chi$ and $z$, we obtain
\[
\langle \frac{dI}{d\omega} \rangle \simeq c_1\frac{27}{14} \cdot 3^{2/3}
\sqrt{\pi} \cdot \Gamma\left(\frac{7}{6}\right)\frac{m^4 c^6 |E_y|}{e^{4/3}
B^{7/3}} p_m^{4/3} \omega^{1/3} \, \mbox{as} \, \omega \to 0.
\]
The asymptotic behavior of $\langle dI/d\omega \rangle$ changes from
$\omega^{1/3}$ to $\omega^{-2}$ somewhere in the region $\omega <
\omega_*$. An approximate value of the frequency $\omega = \omega_0$
when this change occurs is given by
\[
\omega_0 \simeq 0.8 \frac{e B}{mc} \left(\frac{p_m}{mc}\right)^2.
\]
Thus we see that the radiation spectrum of the accelerated particles of the
whole sheet is maximal at $\omega \simeq\omega_0$ with power gain for
$\omega < \omega_0$ and power damping for $\omega > \omega_0$. The
exponential damping occurs for the large frequencies $\omega > \omega_*$.

\begin{figure}
%\centering
\includegraphics[width=84mm]{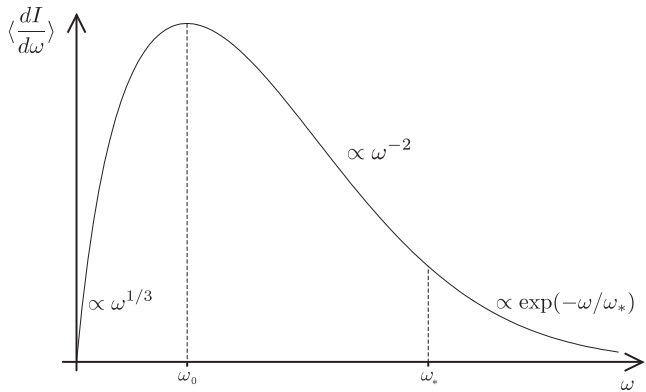}
\caption{Radiation spectrum}
\end{figure}

\section[]{Summary}

According to the papers of~\cite{IK03}, the typical moderate values for the
sheet are:

the typical velocity of the fast magnetic sound is $c_s \simeq 10^9$ cm/s,

the thickness of the sheet is $L \simeq 10^8$ cm,

the typical value of the magnetic field is $B \simeq 10^6$ G,

the plasma temperature in the sheet is $T_0 \simeq 10$ keV.

The time of the sheet compression can be estimated from
Equation~(\ref{etadef}) and the condition $z_+ \sim L$
\[
\Delta t \sim \frac{f_0}{\alpha} \sim c_s / \left|\frac{\partial v}
{\partial t}\right|_{t=0} \sim \frac{L}{c_s} \sim 0.1 \mbox{sec}.
\]
Now we can estimate the maximum value of the Lorentz factor for the
particles ($k=-1/2$):
\[
\Gamma_m = \frac{p_m}{mc}=\left(\frac{T_0}{mc^2}\right)^{3/5}
\left(\frac{L\omega_{c0}}{c_s}\right)^{2/5} \simeq 10^4.
\]
The collimation $\Delta \theta$ of the radiation is determined by the
parameter $K$, $\Delta \theta \simeq K^{-1/2}$,
\[
K=\frac{2}{3} \frac{\Gamma_m \omega_{c0}^2 L r_e}{c^2} \frac{B}{|E_y|},
\]
$r_e$ is the classical electron radius, $r_e = 3 \cdot 10^{-13}$ cm.
Assuming $|E_y|/B \simeq c_s/c$, we obtain $K \simeq 10^7$, which yields
the collimation angle, $\Delta \theta \simeq 3\cdot 10^{-4}$. The intensity
of the accelerated particle radiation in the sheet is maximal at the
frequency $\omega_0 \simeq 8\cdot 10^{20} s^{-1}$, which corresponds
to the energies of the order of $500$ keV.
Asymptotic behavior of the spectrum as well as the frequency $\omega_0$
agree with the observed values of GRBs~\cite{P99}.

Thus, we see that the more precise calculations for the model of
cosmological GRBs proposed by~\cite{IK03} confirm
their estimations. Evolution of the current sheet, particle
acceleration due to the magnetic energy release, and the synchrotron radiation
turn out to agree with what was estimated. The total radiated energy corresponds
to the release of the magnetic energy stored in the sheet ($10^{37}-10^{40} erg$,
see the papers of~\cite{IK03}).
The equivalent radiated energy in the isotropic models is $(3\cdot 10^3)^2$
times greater, i.e. about $10^{44}-10^{47} erg$. The latter energy can be
even greater for closer binary systems (where the distance between a supernova
precursor and the star-companion is less than $10^{13} cm$), and it may
reach values of $10^{50}-10^{53} erg$.
The proposed model explains the main characteristics of observed GRBs.

\section*{Acknowledgements}
This work was supported by the Russian Foundation for Fundamental Research
 (grant number 02-02-16752) and the President of Russian Federation Grant (number NSH-1603.2003.2).

\label{lastpage}


\begin{thebibliography}{20}

\bibitem{P01}
Paczynski, B. 2001, preprint, astro-ph/0103384

\bibitem{P99}
Postnov, K.~A. 1999, Physics-Uspechi, 169, 545

\bibitem{Sakh}
Sakharov, A.~D. 1966,  Sov. Phys. Usp., 9, 294

\bibitem{S65}
Sivukhin, D.~V. 1965,  in Reviews of Plasma Physics,  ed. M.A. Leontovich
(New York: Consultants Bureau), 1, 1

\bibitem{IK03}
Istomin, Ya.~N., Komberg, B.~V. 2002, Astronomy Reports, 46, 1008;
2003, New Astron., 8, 209

\bibitem{LL21}
Landau, L.~D., Lifshitz, E.~M. 1980,  The Classical Theory of Fields
(Course of Theoretical Physics Series, Volume 2) 226

\bibitem{LL22}
Landau, L.~D., Lifshitz, E.~M. 1980,  The Classical Theory of Fields
(Course of Theoretical Physics Series, Volume 2) 215

\bibitem{WW98}
Wang, L.\& Wheeler, J.~C. 1998,  ApJ, 504, L87

\bibitem{RW02}
Reeves, J.~N., Watson, D.\& Osborne, J.~P. 2002, Nat, 416, 512

\bibitem{BIP}
Bereznyak, A.~R., Istomin, Ya.~N. \& Pariev, V.I. 2003, A \& A, 403, 793

\bibitem{LB04}
Lyne, A.~G., Burgay, M.\& Kramer, M. 2004,  preprint, astro-ph/0401086

\end{thebibliography}
\end{document}